\documentclass[12pt]{article}

\usepackage{graphics}

\begin{document}

\hspace{10cm}{\bf NT@UW-03-024}\\

\begin{center}
{\bf \Large Parton distributions in the proton and pion}
\end{center}
\begin{center}
{\bf Mary Alberg$^{a,b}$ and Ernest M. Henley$^{a,c}$}\\

\vspace{0.25 in}
{\small \em $^{a}$ Department of Physics, University of Washington, Seattle,
WA 98195, USA} \\
{\small \em $^{b}$ Department of Physics, Seattle University, Seattle,
WA 98122, USA} \\
{\small \em $^{c}$ Institute for Nuclear Theory, University of Washington, 
Seattle, WA 98195, USA }
\end{center}

\vspace{0.25 in}

\begin{abstract}
We use detailed balance for a hadron composed of quark and gluon Fock 
states to obtain parton distributions in the proton and pion on the basis of 
a simple statistical model.
\end{abstract}

\vspace {0.25 in}
 \section{Introduction: parton distributions in the proton}

There has been considerable interest in the flavor dependence of
the proton's  quark and antiquark distributions. The first measurement
of the $(\bar{d} - \bar{u})$ asymmetry was made by the NMC group \cite{NMC}. The 
integral of this distribution showed a violation of the Gottfried sum
rule. Later, Drell-Yan \cite{NA51,E866} and deep inelastic scattering \cite{Hermes} experiments determined the Bjorken-$x$ dependence of the asymmetry. The meson cloud model 
and the Sullivan process were used to explain the momentum fraction
distribution of $(\bar{d} - \bar{u})$; many other models have been proposed \cite{S&T}.

Most recently, Zhang and collaborators \cite{ZZM,ZZY,ZDM} have used a simple statistical
model to calculate the  $(\bar{d} - \bar{u})$ distribution in the proton. 
They consider the proton to be an ensemble of quark-gluon Fock states, and use detailed balance \cite{ZZM,ZZY} or ``the principle of balance" \cite{ZDM}
to determine the distribution functions for all partons of the proton. Despite its 
simplicity, the model does reasonably well in predicting the distributions
of partons, as well as for that of  $(\bar{d} - \bar{u})$. The excess of
$\bar{d}$ over $\bar{u}$ comes about from the 2:1 ratio of u:d, which 
provides an excess of $u$ quarks for the annihilation of $\bar{u}$'s.

Zhang, Zou, and Yang (ZZY) \cite{ZZY} write a general Fock state expansion for the  
proton as

\begin{equation}
|p> =\sum_{i,j,k} c_{ijk}|\{uud\}\{ijk\}>,
\end{equation}
with $i$ the number of $\bar{u}u$ pairs, $j$ the number 
of $\bar{d}d $ pairs and $k$ the number of gluons. The states are normalized
such that the sum of the probabilities $\rho_{ijk}=|c_{ijk}|^2$ of finding a proton in the state
$|\{uud\}\{ijk\}>$ summed over all $i$, $j$, and $k$ is unity \cite{ZZM},

\begin{equation}
\sum_{i,j,k} \rho_{ijk} = 1 \; .\label{norm}
\end{equation}
In ZZY's statistical model, detailed balance between any two Fock states requires that 

\begin{eqnarray}
\rho_{ijk} N(|\{uud\}\{ijk\}> \rightarrow |\{uud\}\{i^\prime j^\prime k^\prime\}>) 
& \equiv & \nonumber \\
      \rho_{i^\prime j^\prime k^\prime} N(|\{uud\}\{i^\prime j^\prime k^\prime\}> 
\rightarrow |\{uud\}\{ijk\})>,&&
\end{eqnarray}
in which $N(A \rightarrow B))$ is the transfer rate of state $A$ into state $B$. Transfer rates between states are assumed to be proportional to the number of partons that can  split or recombine.  Taking into account two processes, $q \leftrightarrow q\, g$ and
$g \leftrightarrow q\, \bar{q}$, ZZY find that

\begin{equation}
\frac{\rho_{ijk}}{\rho_{000}}=\frac{1}{i!(i+2)!j!(j+1)!k!}\; . \label{ratio}
\end{equation}
This equation, together with the normalization condition (\ref{norm}), determines all the $\rho_{ijk}$. It is clear from this equation that $u\bar{u}$ states, labelled by $i$, are suppressed relative to $d\bar{d}$ states, labelled by $j$.
Summing over all states, Zhang et al. \cite{ZZM} find $(\bar{d} - \bar{u}) \approx 0.124$, remarkably close to the experimental value of $0.118 \pm 0.012$ \cite{E866}.

ZZY determined parton distribution functions for the proton by using a Monte Carlo simulation of the distribution of momenta among the $n$ partons in each Fock state.
The phase space volume $f_n^F$ for $n$ free partons  
is determined by
\begin{equation}
df_n^F = \delta^4 (P-\sum_{i=1}^n p_i) \prod_{i=1}^n \frac{ d^3 p_i
}{ (2\pi)^3 2 E_i}\; ,
\end{equation}
with $P$ and $p_i$ the 4-momenta of the proton and the $i$-th parton, respectively. The masses of the partons are neglected so that $E_i = |\vec{p_i}|$, and
\begin{equation} \label{free}
df_n^F = \delta^4 (P-\sum_{i=1}^n p_i) \prod_{i=1}^n \frac{E_i\, dE_i\, d\Omega_i
}{2 (2\pi)^3}\; .
\end{equation}
ZZY argue that this free parton phase space distribution should be multiplied by $\prod E_i^{-1}$
because partons with smaller momenta spend more time at the center of the
proton where they are almost free; these partons are thus weighted with a higher probability. Then $df_n$, the distribution for confined partons, is:

\begin{equation} 
df_n = \delta^4 (P-\sum_{i=1}^n p_i) \prod_{i=1}^n \frac{ dE_i\, d\Omega_i
}{2 (2\pi)^3}\; .
\end{equation}
We have found that the effect of the weighting factor is quite small, except for the very lowest and highest parton momenta.

From the Monte Carlo distribution of parton momenta $\vec{p_i}$, the parton distributions can be found in terms of the
 light cone variable Bjorken $x$,

\begin{equation}
x_i = \frac{E_i - p_{z\,i}}{M} \;,
\end{equation}
in which $M$ is the proton mass.
 We used RAMBO \cite{RAMBO} for our Monte Carlo event generator. 
Then for an $n$-parton state, for which
$n=3+2(i+j)+k$,
 the $x$-distributions for $\bar{u}$ and $\bar{d}$ are 

\begin {eqnarray}
\bar{u}_{ijk}(x) = f_n(x) i\;   , &
\bar{d}_{ijk}(x) = f_n(x) j\;   ,
\end {eqnarray}
 for $u$ and $d$ are
\begin {eqnarray}
u_{ijk}(x) = f_n(x) (2 + i)\;   , &
d_{ijk}(x) = f_n(x) (1+ j)  \;  ,
\end {eqnarray}
and for the gluons is
\begin {equation}
g_{ijk}(x) = f_n(x)  k  \;  .
\end {equation}
Thus, we find, in accord with ZZY, 
\begin{equation}
\bar{u}(x) = \sum_{i,j,k} \rho_{ijk}\, \bar{u}_{ijk}(x) \; ,\label{pdf}
\end{equation}
and corresponding equations for $\bar{d}(x)$, $u(x)$, $d(x)$ and $g(x)$,
normalized so that
\begin{equation}
\int_0^1 x  [u(x) + d(x)+\bar{u}(x)+\bar{d}(x) +g (x)] dx = 1 \; .
\end{equation}
The average number of partons in the proton, $\bar{n}$, is given by
\begin{equation}
\bar{n}=\int_0^1 [u(x) + d(x)+\bar{u}(x)+\bar{d}(x) +g (x)] dx \approx 5.6\; .
\end{equation}
This sets the scale $\mu_0 \approx \bar{E}=M/\bar{n}\approx 0.17$ GeV for $Q_0^2=\mu_0^2$ at which the distributions are calculated.
The plot in Fig. 1, which reproduces ZZY's results, shows that the experimentally deduced 
($\bar{d} - \bar{u}$) is fit qualitatively in
this model , but is low at small $x$ and high at large $x$. The discrepancy with experiment shows 
up more starkly in Fig. 2, our plot of $\bar{d}(x)/\bar{u}(x)$. We find that these results are changed very little if the phase space weighting factors $\prod E_i^{-m}$ are varied from $ m=1$ to 
 $m=3$.
 Nevertheless, we believe that 
it is remarkable that such a simple model does so well.

\begin{figure}
\vspace{30mm} % height of figure
\centerline{\includegraphics{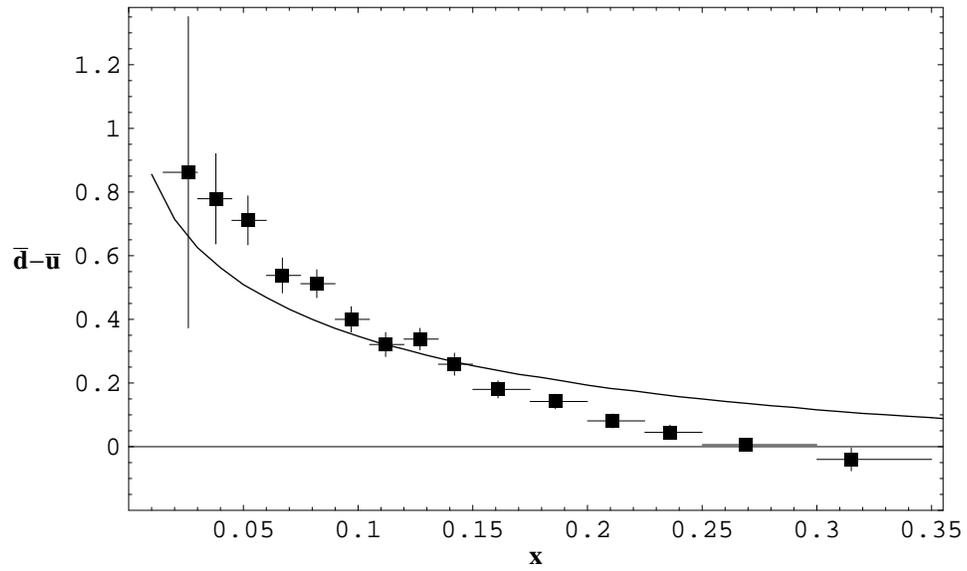}}
\caption{Comparison of statistical model calculation with E866 experimental results \cite{E866} for $\bar{d}-\bar{u}$.}
\end{figure}

\begin{figure}
\vspace{30mm} % height of figure
\centerline{\includegraphics{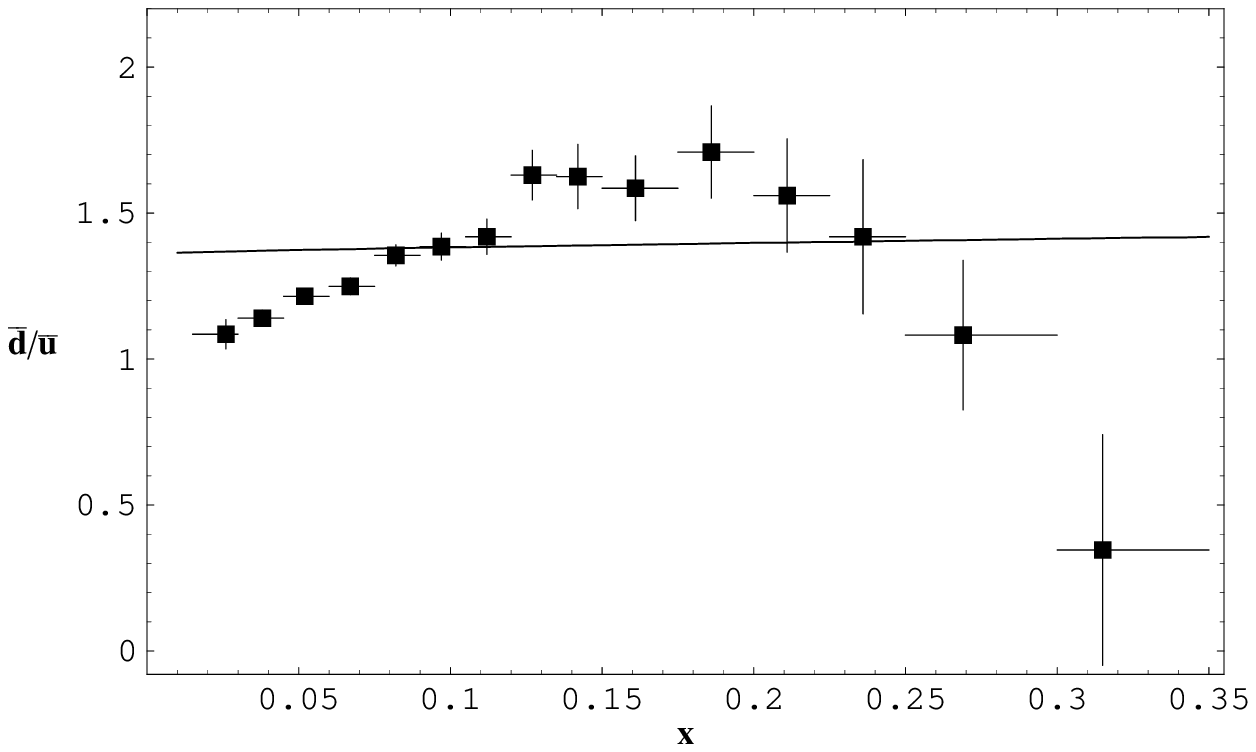}}
\caption{Comparison of statistical model calculation with E866 experimental results \cite{E866} for  $\bar{d}/\bar{u}$.}
\end{figure}

There are, of course, other explanations of the excess of $\bar{d}$ over
$\bar{u}$, which have been considered previously, particularly the pion cloud 
\cite{Kumano:1991mj,Signal:1991ug,Koepf:1996yh,MST,NSSS} and $\omega$ mesons \cite{AHM}.  The leading term in the Fock state expansion of the pion cloud model is  a ``bare" proton which consists of valence quarks 
plus $\bar{q}q$ pairs due to gluon splitting. It can be argued that
this gluon splitting is what is being considered by ZZY and thus the perturbative sea in the
``bare" proton should not be symmetric, i.e. $\bar{d} \ne \bar{u}$. The pion cloud, represented by the higher-order terms in the expansion, would then be an 
additional effect. If this is the case, the pions must have
a larger role at high $x$
where the fall-off of the ratio $\bar{d}/\bar{u}$ is not reproduced by the 
statistical model. We will not pursue this argument further here. 

\section{Parton distributions in the pion}

 If the statistical model has some validity, then it should not only work for
 the proton, but also for the pion. This distribution function is, in fact,
 required  in the pion cloud model. We have therefore investigated the valence
 and sea quark distributions for the $\pi^+$ in the statistical model. The
 formulas are similar to those for the proton, but because 
 there is only one valence quark of each flavor in the $\pi^+$, the sea is flavor symmetric. 

We write the Fock state expansion for the pion as

\begin{equation}
|\pi^+> =\sum_{i,j,k} c_{ijk}|\{u\bar{d}\}\{ijk\}>.
\end{equation}
The analysis of section 1 is unchanged, except that  
 $n=2+2(i+j)+k$,
and the ratio of probabilities for different Fock states (\ref{ratio})
is now

\begin{equation}
\frac{\rho_{ijk}}{\rho_{000}}=\frac{1}{i!(i+1)!j!(j+1)!k!}\; ,
\end{equation}
so that the $\pi^+$ sea is symmetric, i.e. $\bar{u}(x)=d(x)$. We find $\bar{n}= 4.5$ in the pion.

 The quark distributions for a specific $n$-parton state are
\begin {eqnarray}
\bar{d}_{ijk}(x) = f_n(x) (1+j)\;   , &
u_{ijk}(x) = f_n(x) (1 + i) \; ,
\end {eqnarray}
\begin {eqnarray}
\bar{u}_{ijk}(x) = f_n(x) i\;   , &
d_{ijk}(x) = f_n(x)  j  \;  ,
\end {eqnarray}
and for the gluons is
\begin {equation}
g_{ijk}(x) = f_n(x)  k  \;  .
\end {equation}
The parton distributions are found by summing these distribution functions over all values of $\{ijk\}$
\begin{equation}
u(x) = \sum_{i,j,k} \rho_{ijk}\, u_{ijk}(x)= \bar{d}(x) \; ,
\end{equation}
\begin{equation}
d(x) = \sum_{i,j,k} \rho_{ijk}\, d_{ijk}(x)= \bar{u}(x) \; ,
\end{equation}
and
\begin{equation}
g(x) = \sum_{i,j,k} \rho_{ijk}\, g_{ijk}(x)\; .
\end{equation}
The valence quark distribution function is
\begin{equation}
v(x) = u(x)-\bar{u}(x)=\bar{d}(x)-d(x)
\end{equation}
\begin{figure}
\vspace{30mm} % height of figure
\centerline{\includegraphics{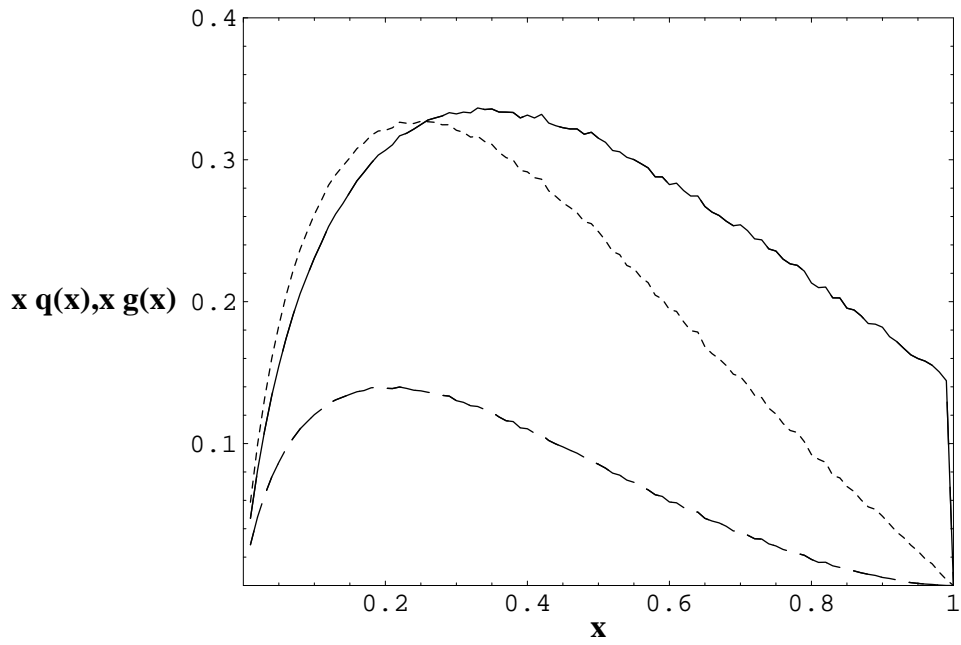}}
\caption{Our results for parton density distributions $x\, q(x)$ and $x\, g(x)$  for the pion. Solid curve: valence quark distribution; long-dashed curve: sea quark distribution; short-dashed curve: gluon distribution.}
\end{figure} 
Our results are shown in Fig. 3.  The valence quark distributions 
 are too high for large $x$ because of the dominant contribution of the $n=2$, $\{ijk\}=\{000\}$ state, the leading term in the Fock expansion, for which 
$f_{2}(x)$ is a uniform distribution in $x$. The sea quark distribution is flavor symmetric, as noted above.
\begin{figure}
\vspace{30mm} % height of figure
\centerline{\includegraphics{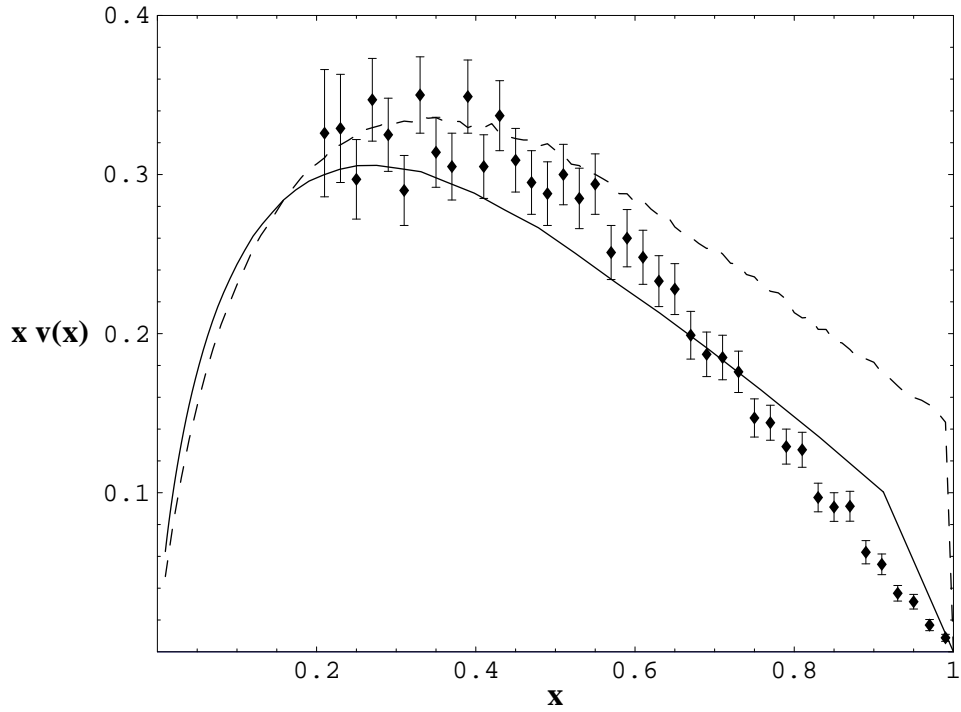}}
\caption{Our calculation of the valence quark distribution $x\, v(x)$ in the pion, compared to the experimental results of Conway et al. \cite{Conway}. The dashed curve shows our results without any evolution, as in Fig. 3, which correspond to a scale of $Q_0^2 = 1.96$ GeV$^2$.
The solid curve shows our results evolved to $Q^2=16$ GeV$^2$ of the E615 experiment.}
\end{figure}
In order to compare our valence quark distributions to experiment, we carried out an evolution in $Q^2$.  We determined the starting scale of our distributions by requiring that  the first and second moments of our valence quark distribution at $Q^2 = 4$ GeV$^2$ be equal to those found by Sutton et al. \cite{Sutton}. This gave us a starting scale of $Q_0^2 = 1.96$ GeV$^2$. We used Miyama and Kumano's code BF1 \cite{Miyama} for the DGLAP \cite{DGLAP} evolution.
We compare our pion valence quark distribution with
that obtained by E615 from pion scattering on tungsten \cite{Conway} in Fig. 4. The dashed curve shows our results without any evolution, as in Fig. 3.
The solid curve shows our results evolved to $Q^2=16$ GeV$^2$ of the E615 experiment. The agreement between theory and experiment is good.
\begin{figure}
\vspace{30mm} % height of figure
\centerline{\includegraphics{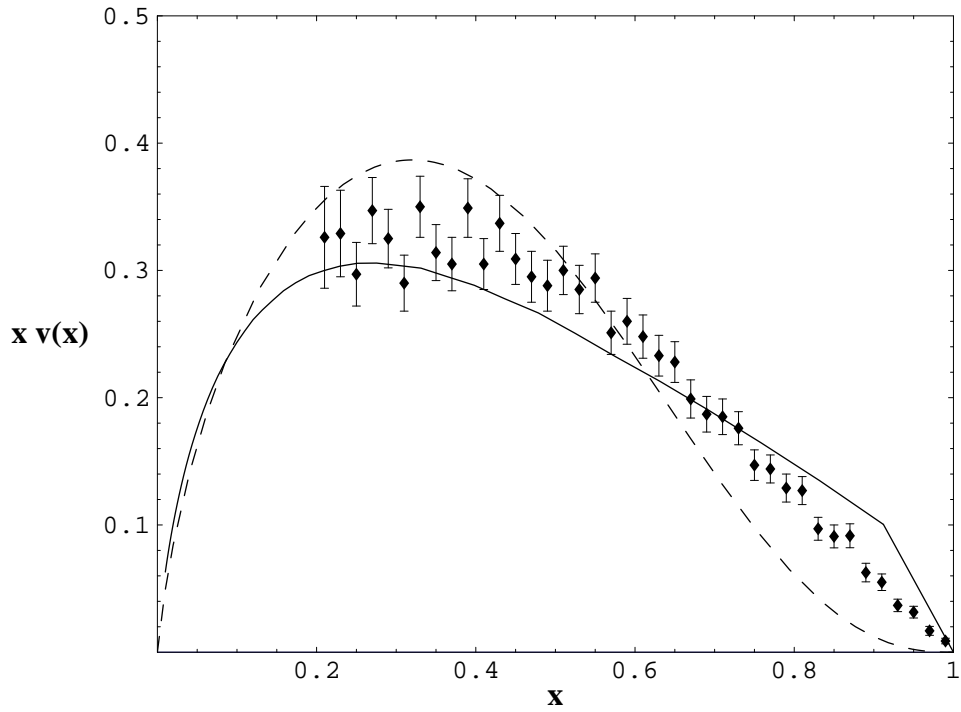}}
\caption{Our results (solid curve) for the valence quark distribution $x\, v(x)$ in the pion, compared to the calculation of Hecht, Roberts and Smith \cite{Hecht} (dashed curve) and the experimental results of Conway et al. \cite{Conway}. Both calculations were evolved to $Q^2=16$ GeV$^2$ of the E615 experiment.}
\end{figure}
Other theoretical calculations of pion parton distribution functions have used constituent quark models, the Nambu-Jona-Lasinio model, instantons, or the Dyson Schwinger equations. Moments of the distributions can be calculated in lattice QCD, from which particular forms of the distributions can be reconstructed. For references see the recent papers of Hecht, Roberts and Schmidt \cite{Hecht} and Detmold, Melnitchouk and Thomas \cite{Detmold}. In Fig. 5 we compare our valence quark distribution to the Dyson-Schwinger calculation of Hecht et al. and to experiment. Both distributions were evolved to $Q^2=16$ GeV$^2$. We find it remarkable that our simple statistical model agrees with experiment as well as  the covariant, QCD-based model.

\section{Conclusions}

The calculation of parton distribution functions is an important goal of non-perturbative QCD. We have used the statistical model of Zhang et al., developed for the calculation of parton distribution functions in the proton, to calculate the parton distribution functions of the pion. We find that this simple model, with no free parameters, is in good agreement with experiment and other calculations.                                                                           
                                                                               
\section{Acknowledgement}

We thank Toby Burnett and Steve Ellis for discussions on the use of RAMBO, and Michael Clement for computational work with RAMBO and calculation of distribution functions.
This work was supported in part by the U.S. Department of Energy and the Research in Undergraduate Institutions program of the U.S. National Science Foundation, Grant No.0070942.

\end{document}